\documentclass[aps,pre,superscriptaddress,showpacs]{revtex4}
\usepackage{graphics}
\usepackage{graphicx}
\usepackage{epsfig}
\usepackage{amsmath}
\usepackage{amssymb}

\newcommand{\nd}{n}

\newcommand{\Sc}{\ensuremath{{\rm Sc}}}

\newcommand{\bc}{\mbox{\boldmath$v$}}

\newcommand{\bTheta}{\mbox{\boldmath$\Theta$}}

\newcommand{\jj}{{\mbox{\boldmath$j$}}}
\newcommand{\qq}{{\mbox{\boldmath$q$}}}

\newcommand{\uu}{\mbox{\boldmath$u$}}

\begin{document}
\title{Quasi-Equilibrium Lattice Boltzmann Method}
\author{S. Ansumali}
\affiliation{CBE, Nanyang  Technological University, 639798
Singapore, Singapore }

\author{S. Arcidiacono}
\affiliation{Paul Scherrer Institute, Combustion Research Unit,
5232 Villigen PSI, Switzerland }

\author{S. S. Chikatamarla}
\affiliation{ETH Z\"urich, Aerothermochemistry and Combustion
Systems Lab, CH-8092 Z\"urich, Switzerland}

\author{ N. I. Prasianakis}
\affiliation{ETH Z\"urich, Aerothermochemistry and Combustion
Systems Lab, CH-8092 Z\"urich, Switzerland}

\author{A. N. Gorban}
\affiliation{Department of Mathematics, University of Leicester,
Leicester LE1 7RH, UK}

\author{I. V. Karlin\footnote{Corresponding author}}
\affiliation{ETH Z\"urich, Aerothermochemistry and Combustion
Systems Lab, CH-8092 Z\"urich, Switzerland}
%

\begin{abstract} A general lattice Boltzmann method for simulation
of fluids with tailored transport coefficients is presented. It is
based on the recently introduced quasi-equilibrium kinetic models,
and a general lattice Boltzmann implementation is developed.
Lattice Boltzmann models for isothermal binary mixtures with a
given Schmidt number, and for a weakly compressible flow with a
given Prandtl number are derived and validated.
\end{abstract}
 \maketitle

%
\section{Introduction}
\label{intro}

The lattice Boltzmann (LB) method is a powerful approach to
hydrodynamics, with applications ranging from large Reynolds
number flows to flows at a micron scale, porous media and
multi-phase flows \cite{succi}. However, in spite of its rapid
development over a decade, the method is still lacking a
systematic and flexible construction of numerical schemes for
situations beyond a simple fluid, such as mixture models with a
given Schmidt number, or thermal models with a given Prandtl
number.

 In this paper, we introduce a general method of
constructing the lattice Boltzmann models. Our approach is based
on recently introduced quasi-equilibrium (QE) kinetic models
\cite{GK94,Levermore}. The structure of the paper is as follows:
First, for the sake of completeness, we remind the construction of
continuous time-space QE models. After that, we derive a lattice
Boltzmann discretization scheme of the QE models. The resulting
quasi-equilibrium lattice Boltzmann method (QELBM) is illustrated
with two examples of particular interest. The first example is the
thermal model with a prescribed Prandtl numbers. The second
example is the isothermal binary mixture model with a prescribed
Schmidt number. Basic steps of construction of these models are
presented and numerical validation is provided.


\section{Quasi-Equilibrium Kinetic Models}

\subsection{General}
We address the class of lattice Boltzmann models equipped with the
Boltzmann entropy function of the form,
\begin{equation}
H=\sum_{i=1}^{n} f_i \ln \left(\frac{f_i}{W_i}\right).
\label{HfunctionBoltzmann}
\end{equation}
where $f_i\ge0$ are populations of the discrete velocities $v_i$,
$i=1,\dots,n$, and $W_i>0$ are corresponding weights. A wide class
of relevant entropy functions (\ref{HfunctionBoltzmann}) pertinent
to simulation of hydrodynamics was described in
\cite{Karlin99,AK5,AK_PRL_05,CAK_PRL_06,CK_PRL_06}. In the
examples below we shall use the $H$-functions for the isothermal
models \cite{Karlin99} and the recently introduced weakly
compressible thermal models \cite{AK_PRL_05} in two dimensions,
for which the set of nine velocities $v_i$ and corresponding
weights $W_i$ are
\begin{align}
\label{velset}
\begin{split}
v_x &= \left\{0, 1, 0, -1, 0, 1,-1,-1,1\right\}\\
v_y &= \left\{0, 0, 1,  0, -1, 1,1,-1,-1\right\}\\ W &=(1/36)
\left\{16, 4, 4,  4, 4, 1,1,1,1\right\}.
\end{split}
\end{align}

 We begin with a generic
construction of the quasi-equilibrium kinetic models
\cite{GK94,Levermore} specified for the discrete velocity case.
Let $M=\{M_1,\dots,M_{k_M}\}$ be a set of locally conserved
fields, and $N=\{N_1,\dots,N_{k_N}\}$ be a set of quasi-conserved
slow fields. Functions $M_m$ and $N_l$ are assumed linear
functions (moments) of the populations, $M_m=\sum_{i}M_{mi}f_i$
and $N_l=\sum_{i}N_{li}f_i$. The choice of $M$ and $N$ depends on
the particular problem but we always assume that the density is
included into the list $M$, $\rho=M_1=\sum_{i=1}^nf_i$. The
equilibrium population vector $f^{\rm eq}(M)$ is defined as the
minimizer of the $H$-function (\ref{HfunctionBoltzmann}) under
fixed $M$. The quasi-equilibrium $f^*(M,N)$ is defined as the
minimizer of $H$ under fixed $M$ and $N$. By construction,
functions $f^*$ and $f^{\rm eq}$ satisfy consistency relations
\begin{eqnarray}
 M(f^*(M,N))&=&M,\nonumber\\  N(f^*(M,N))&=&N,\nonumber\\ M(f^{\rm
eq}(M))&=&M.\label{consist}
\end{eqnarray}
The quasi-equilibrium kinetic model reads

\begin{equation}
\partial_t f_i+ v_{i\alpha} \partial_{\alpha} f_i =
-\frac{1}{\tau_{1}}\left[f_i-f_i^{*}(M(f),N(f))\right]-
\frac{1}{\tau_{2}}\left[f_i^{*}(M(f),N(f))-f_i^{\rm eq}(M(f))
\right].\label{QEM}
\end{equation}
Denoting the right hand side (the collision integral) as $Q_i$, it
is easy to see that consistency condition implies local
conservation laws, $M(Q)=0$. Note that the first part of the
collision integral which describes relaxation to the
quasi-equilibrium, also conserves the $N$-fields, $N(f-f^*)=0$.
Moreover, it is straightforward to prove the $H$-theorem. For
that, it suffices to rewrite
\begin{equation}
Q_i=-\frac{1}{\tau_{2}}\left[f_i-f_i^{\rm
eq}\right]-\left(\frac{\tau_2-\tau_1}{\tau_1\tau_2}\right)\left[f_i-f_i^{*}\right].
\end{equation}
The entropy production $\sigma=\sum_{i=1}^n(\partial H/\partial
f_i)Q_i$ becomes
\begin{equation}
\sigma=-\frac{1}{\tau_2}\sum_{i=1}^n\ln\left(\frac{f_i}{f_i^{\rm
eq}}\right)(f_i-f_i^{\rm
eq})-\left(\frac{\tau_2-\tau_1}{\tau_1\tau_2}\right)
\sum_{i=1}^n\ln\left(\frac{f_i}{f_i^*}\right) (f_i-f_i^*),
\end{equation}
and is non-positive semi-definite {\it provided} relaxation times
satisfy the hierarchy,
\begin{equation}
\label{hierarchy} \tau_1\le\tau_2.
\end{equation}
Thus, in the QE models, relaxation to the equilibrium is split in
two steps. In the first step, the distribution functions relaxes
to the quasi-equilibrium with the faster relaxation time $\tau_1$.
In the second step, the quasi-equilibrium  relaxes to the
equilibrium with the slower relaxation times $\tau_2$. If
$\tau_1=\tau_2$, the intermediate relaxation step to the
quasi-equilibrium disappears from (\ref{QEM}), and it reduces to
the Bhatnagar-Gross-Krook (BGK) model.

\subsection{Triangle entropy method}
For a practical implementation, explicit form of the functions
$f_i^{\rm eq}(M)$ and $f_i^*(M,N)$ are required. While for most of
the cases, the equilibrium $f_i^{\rm eq}$ can be found explicitly
either in a closed form or in a form of expansion, explicit
construction of the quasi-equilibrium is case-dependent. Here we
suggest a simple way to find quasi-equilibria in explicit form by
perturbation around the equilibrium \cite{GK94,GK06}. Let us
assume that the equilibrium $f^{\rm eq}(M)$ has been found
explicitly, and that near the equilibrium
\begin{equation}
\label{triang} f^*=f^{\rm eq}(M)+\delta f^*(M,N).
\end{equation}
Expanding the entropy function $H$ (\ref{HfunctionBoltzmann}) at
the equilibrium to quadratic terms, we obtain
\begin{equation}
 H(f^{\rm eq}+\delta f)=H(f^{\rm
eq})+\sum_{i=1}^n\delta f_i\left(\ln\left(\frac{f_i^{\rm
eq}}{W_i}\right)+1\right)+\frac{1}{2}\sum_{i=1}^n\frac{\delta
f_i^2}{f_i^{\rm eq}}+O(\delta f^3).\label{DeltaH}
\end{equation}
The quadratic expansion of the entropy function (\ref{DeltaH})
maintains convexity, and we find its minimizer $\delta f^*$
subject to the linear constraints
\begin{eqnarray}
 \sum_{i=1}^n M_{ki}\delta f_i^*&=&0,\nonumber\\
\sum_{i=1}^n N_{li}\delta f_i^*&=&N_l-N_l^{\rm
eq},\label{TriangleConstraints}
\end{eqnarray}
where $N^{\rm eq}=N(f^{\rm eq}(M))$ are the values of the
non-conserved fields at equilibrium. Solution to this minimization
problem has the form,
\begin{equation}
 f^*=f_i^{\rm eq}(M)+\delta f^*_i=f_i^{\rm
eq}(M)\left(1+\sum_{s=1}^{k_M}\lambda_s
M_{si}+\sum_{s=1}^{k_N}\chi_s N_{si}\right),\label{TriangleQE}
\end{equation}
where Lagrange multipliers $\lambda_s$ and $\chi_s$ are found upon
substituting function (\ref{TriangleQE}) into the constrains
(\ref{TriangleConstraints}), and solving the $(k_M+k_N)\times
(k_M+k_N)$ linear algebraic problem.  Introducing a
$(k_M+k_N)$-dimensional vector of Lagrange multipliers,
$(\lambda,\chi)=(\lambda_1,\dots,\lambda_{k_M},\chi_1,\dots,\chi_{k_N})$,
and matrices
\begin{eqnarray}
\left(A_{MM}\right)_{kl}&=&\sum_{i=1}^n M_{ki}f_i^{\rm
eq}M_{li},\ k,l=1,\dots,k_M\nonumber\\
 \left(A_{MN}\right)_{kl}&=&\sum_{i=1}^n
M_{ki}f_i^{\rm eq}N_{li}, \ k=1,\dots,k_{M},\ l=1,\dots,k_N\nonumber\\
\left(A_{NN}\right)_{kl}&=&\sum_{i=1}^n N_{ki}f_i^{\rm eq}N_{li},
\ k,l=1,\dots,k_N,
\end{eqnarray}
we find the solution in the matrix form,
\begin{equation}
\left(%
\begin{array}{c}
  \lambda \\
  \chi \\
\end{array}%
\right)=\left(%
\begin{array}{cc}
  A_{MM} & A_{MN} \\
  A_{MN}^T & A_{NN} \\
\end{array}%
\right)^{-1}\left(%
\begin{array}{c}
  0 \\
  N-N^{\rm eq} \\
\end{array}%
\right), \label{TriangleSolution}
\end{equation}
where $T$ denotes transposition, and $N-N^{\rm eq}$ is the
$k_N$-dimensional vector $(N_1-N_1^{\rm eq},\dots,
N_{k_N}-N_{k_N}^{\rm eq})$. Note that the solution depends
linearly on the deviation of the non-conserved fields $N-N^{\rm
eq}$ and non-linearly on the locally conserved $M$. For this
quasi-linear quasi-equilibrium, the entropy production becomes
\begin{equation}
\sigma=-\frac{1}{\tau_2}\sum_{i=1}^n\frac{(f_i-f_i^{\rm
eq})^2}{f_i^{\rm
eq}}-\left(\frac{\tau_2-\tau_1}{\tau_1\tau_2}\right)
\sum_{i=1}^n\frac{(f_i-f_i^*)^2}{f_i^{\rm eq}} +O(\delta
f^3).\label{TriangleSigma}
\end{equation}
Thus, with the use of the triangle entropy method, the kinetic
model satisfies the entropy production inequality (both of the two
quadratic forms in (\ref{TriangleSigma}) are non-positive
semi-definite) once the populations remain close to the local
equilibrium. This is sufficient for most of the applications.

\section{Quasi-Equilibrium Lattice Boltzmann Method} We shall now
derive a second-order time discretization for the generic kinetic
equation (\ref{QEM}). Since the derivation only uses the
consistency condition (\ref{consist}), it is equally applicable to
exact quasi-equilibria and to those obtained by the triangle
entropy method (\ref{triang}). Following \cite{He}, kinetic
equations (\ref{QEM}) are integrated in time from $t$ to $t+\delta
t$ along characteristics, and the time integral of the right hand
side is evaluated by trapezoidal rule. Introducing a map
\begin{equation}
\label{Map}
 f_i\to g_i=f_i-\frac{\delta t}{2}Q_i(f),
\end{equation}
the result is written as
\begin{equation}
 g_i(x+c_i\delta t,t+\delta
t)=g_i(x,t)-\omega_1[g_i(x,t)-f_i^*(x,t)]-\frac{\omega_1\tau_1}{\tau_2}[f_i^*(x,t)-f_i^{\rm
eq}(x,t)],\label{SecondOrder1}
\end{equation}
where
\begin{eqnarray}
\omega_1&=&\frac{2\delta t}{2\tau_1+\delta t},\nonumber\\
f^*(x,t)&=&f^*(M(f),N(f)),\nonumber\\
f_i^{\rm eq}(x,t)&=&f_i^{\rm eq}(M(f)).
\end{eqnarray}
Note that the right hand side in (\ref{SecondOrder1}) still
contains terms which depend on populations $f$ rather than on the
function $g$ (\ref{Map}).
In order to obtain a closed-form equation solely in terms of
functions $g$, we first remark that, taking locally conserved
moments of the map (\ref{Map}) we have (same as in the BGK case)
$M(f)=M(g)$, thus $f^{\rm eq}(f)=f^{\rm eq}(g)$ in the second term
in (\ref{SecondOrder1}). For the quasi-conserved fields $N$, the
situation is slightly different. Evaluating the moments $N$ of the
map (\ref{Map}), we obtain
\begin{equation}
N(g)=N(f)+\frac{\delta t}{2\tau_2}(N(f)-N^{\rm eq}(f)).
\end{equation}
Inverting this relation, and substituting it into
(\ref{SecondOrder1}), after simple transformations we  obtain:
%

\begin{equation}
 g_i(x+c_i\delta t,t+\delta
t)=(1-\omega_1)g_i+\omega_1\left(\frac{\tau_1}{\tau_2}\right)f_i^{\rm
eq}(M)+\omega_1\left(\frac{\tau_2-\tau_1}{\tau_2}\right)
f_i^*(M,N'),\label{deliverable}
\end{equation}
where $M=M(g)$, while $N'$ is evaluated according to the rule
\begin{equation}
N'=\left(1-\frac{\delta t}{2\tau_2+\delta
t}\right)N(g)+\frac{\delta t}{2\tau_2+\delta t}N^{\rm eq}(g).
\end{equation}
Equation (\ref{deliverable}) is the basic second-order time
stepping algorithm for the quasi-equilibrium lattice Boltzmann
models, and is the main result of this paper. It is important to
note the shift in the dependence of the quasi-equilibrium
population $f^*$, it does not depend just on $N(g)$ but rather on
a convex linear combination between $N(g)$ and the equilibrium
value $N^{\rm eq}(g)$. Discretization in space depends on the
problem at hand (in the simplest case, the lattice Boltzmann
discretization is readily applicable). We shall now proceed with
specific examples of QELBM.

\section{Examples}
\subsection{One-component fluid with a given Prandtl number}
Navier-Stokes equations for a one-component compressible fluid are
characterized by the Prandtl number, ${\rm Pr}=(C_p\mu)/\kappa$,
where $C_p$ is the specific heat at constant pressure
($C_p=(D+2)/2$ is specific heat of ideal gas below), $\mu$ is
viscosity coefficient, and $\kappa$ is thermal conductivity. When
a single relaxation time kinetic equation is used (for example,
the BGK model), this results in a fixed Prandtl number. As we
shall see it below, physical consistency of the QE models, the
relaxation time hierarchy (\ref{hierarchy}), implies that two
models with two different quasi-equilibria are required in order
to cover the entire range of Prandtl number. For the sake of
concreteness, we consider a simple, weakly compressible flow model
on the standard two-dimensional $D2Q9$ lattice
\cite{AK_PRL_05,KyotoThermal}. The locally conserved fields $M$
are density $\rho$, momentum density $\jj$, and pressure density
$p$,

\begin{equation}\label{FieldsHydro}
\sum_{i=1}^{\nd} \{ 1,\, \bc_{i}, v_i^2\} f_i =\{\rho, \jj, 2p+
\rho^{-1}j^2\}.
\end{equation}
To second order in the momentum, the equilibrium reads
\cite{AK_PRL_05,KyotoThermal}:
\begin{equation}
f_i^{\rm eq} (\rho ,\jj,p) = \rho \left(1 -
\frac{p}{\rho}\right)^2 \left(\frac{\frac{p}{\rho}}{{2(1 -
\frac{p}{\rho} )}}\right)^{v_i^2 } \left[1 + \frac{{v_{i\alpha}
j_{\alpha} }}{p} + \frac{{j_{\alpha}j_{\beta} }}{{2p^2
}}\left(v_{i\alpha}v_{i\beta } - \frac{{4(\frac{p}{\rho})^2  +
v_i^2 (1 - 3\frac{p}{\rho} )}}{{2(1 - \frac{p}{\rho} )}}\delta
_{\alpha\beta } \right)\right].
\label{Tequilibrium}
\end{equation}
This model operates in a small temperature window $T=p/\rho$,
around the reference temperature $T_0=1/3$ (see Refs.\
\cite{AK_PRL_05,KyotoThermal} for an estimate of the temperature
window). We remark that, due to a low symmetry of the velocity
set, the lattice BGK (LBGK) model with the equilibrium
(\ref{Tequilibrium}) gives ${\rm Pr}_{\rm LBGK}=4$
\cite{KyotoThermal} (not ${\rm Pr}=1$, as in the continuous
kinetic theory or thermal lattice Boltzmann models with a higher
symmetry \cite{AK5,CK_PRL_06}).
In the first case (${\rm Pr}<{\rm Pr}_{\rm LBGK}$), the specified
slow variables $N$ are the components of the  heat flux $\qq$,
\begin{equation}
 q_{\alpha} = \sum_{i=1}^{\nd}
 \left(v_{i\alpha}-u_{\alpha}\right)\left( \bc_{i}-
 \uu\right)^2f_i.
 \label{Pr<}
\end{equation}
where $\uu=\jj/\rho$ is the mean velocity.
 In the opposite case (${\rm Pr}>{\rm Pr}_{\rm LBGK}$), the slow variables $N$ are the components of the stress
 tensor $\bTheta$
 \begin{equation}
 \Theta_{\alpha\beta} = \sum_{i=1}^{\nd}\left[
 \left(v_{i\alpha}-u_{\alpha}\right)\left(v_{i\beta}-
 u_{\beta}\right)-\frac{2}{D}\delta_{\alpha\beta}( \bc_{i}-
 \uu)^2\right]f_i.
 \label{Pr>}
 \end{equation}
Explicit expressions for the pertinent quasi-equilibria,
$f_i^{*}(\rho, \jj, p, \qq)$ (case (\ref{Pr<}), $\Pr\le\Pr_{\rm
LBGK}$) and $f_i^{*}(\rho, \jj, p, \bTheta)$ (case (\ref{Pr>}),
$\Pr\ge\Pr_{\rm LBGK}$) are easily found using the explicit
formulas of the triangle entropy method, Eqs. (\ref{TriangleQE})
and (\ref{TriangleSolution}), and are not displayed here. Using
the Chapman-Enskog method, we derive the hydrodynamic equations
(Navier-Stokes-Fourier) for the density, momentum and temperature
from the continuous QE models (\ref{QEM}) with the corresponding
quasi-equilibria $f_i^{*}(\rho, \jj, p, \qq)$ and $f_i^{*}(\rho,
\jj, p, \bTheta)$ (note that the explicit form of $f_i^*$ is not
required to perform this analytical computation, the consistency
condition \ref{consist} suffices). The viscosity $\mu$ and the
thermal conductivity $\kappa$ coefficients thus obtained, imply
the following Prandtl number:

\begin{eqnarray}
\mu=\rho T_0\tau_1, \ \kappa=\frac{1}{2}\rho T_0\tau_2\
\Rightarrow\ {\rm Pr}=4\frac{\tau_1}{\tau_2}\le{\rm Pr}_{\rm
LBGK}\quad {\rm \
(\ref{Pr<})},\\
 \mu=\rho T_0\tau_2, \ \kappa=\frac{1}{2}\rho T_0\tau_1\ \Rightarrow\ {\rm Pr}=4\frac{\tau_2}{\tau_1}\ge{\rm Pr}_{\rm LBGK}\quad {\rm
\ (\ref{Pr>})}.
\end{eqnarray}
A comment to these formulas is in order: The hierarchy of
relaxation times ($\tau_1\le\tau_2$) implies a restriction on the
range of admissible Prandtl number when a specified
quasi-equilibrium is used. In the case where the heat flux is
considered as the slow variable, the thermal conductivity
(relaxation rate of the heat flux) is proportional to the slow
relaxation time, $\kappa\sim\tau_2$, while $\mu\sim\tau_1$. In the
opposite case when the stress tensor is chosen as a slow variable,
the dependence is inverted, $\mu\sim \tau_2$ and
$\kappa\sim\tau_1$.

\begin{figure}[htbp]
\centering
  \includegraphics[scale=0.6]{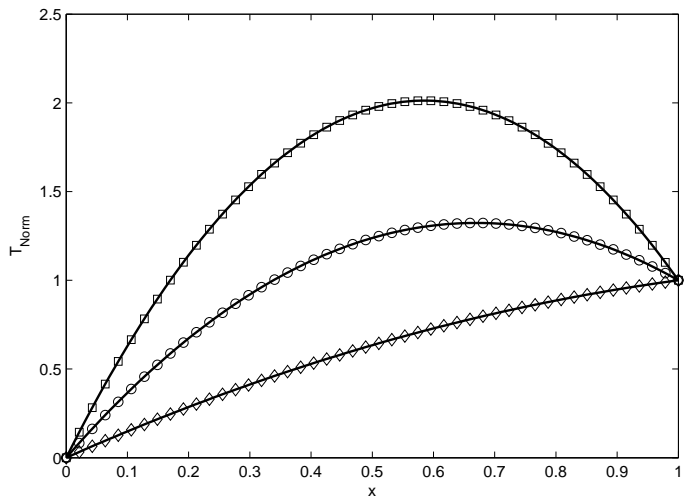}
\caption{Steady state temperature profile in Couette flow between
parallel plates moving with a relative velocity $U$, at the
temperature difference $\delta T$. Lower curve: QELBM (\ref{Pr<})
with $\Pr=0.710$ (air); Middle curve: LBGK model with $\Pr=4$;
Upper curve: QELBM (\ref{Pr>}) with $\Pr=8$. Symbol: simulation;
Line: analytical solution. The Eckert number, ${\rm
Ec}=(U^2)/(C_p\delta T)$ is $1.5$.}
\label{Fig1}
\end{figure}

Implementation of the present QE models is a straightforward
application of equation (\ref{deliverable}). In Fig.\ \ref{Fig1},
simulation of Couette flow between parallel plates at different
temperatures is compared with the analytical solution. Diffusive
boundary condition was used \cite{AK_diffusive_BC}. Agreement
between simulation and analytical solution is excellent.

%

\subsection{Binary mixture at a given Schmidt number}

Our second example is the isothermal binary mixture of ideal
fluids, A and B, with particles masse $m_{\rm A,B}$. The locally
conserved variables $M$ are the densities of the components,
$\rho_{\rm A,B}$, and the momentum of the mixture $\jj=\jj_{\rm
A}+\jj_{\rm B}$, where $\jj_{\rm A,B}$ are the momenta of the
components. The efficiency of diffusion mixing versus viscous
dissipation of momentum in various fluids is characterized by the
Schmidt number, $\Sc=\mu/(\rho D_{\rm AB})$, where $\mu$ is the
viscosity coefficient, $D_{\rm AB}$ is the binary diffusion
coefficient, and $\rho=\rho_{\rm A}+\rho_{\rm B}$. Let us further
introduce the molar fractions $X_{\rm A,B}=n_{\rm A,B}/(n_{\rm
A}+n_{\rm B})$ where $n_{\rm A,B}=\rho_{\rm A,B}/m_{\rm A,B}$ is
the number density of components, and the reference Schmidt
number, $\tilde{\Sc}=m_{\rm AB}/(\rho X_{\rm A}X_{\rm B})$ with
$m_{\rm AB}=(\rho_{\rm A}\rho_{\rm B})/(\rho_{\rm A}+\rho_{\rm
B})$ the reduced mass density. We shall now consider a pair of
quasi-equilibrium models with two different quasi-equilibria which
cover the entire range of Schmidt number,
$\Sc\lessgtr\tilde{\Sc}$. For concreteness, we use the standard
isothermal $D2Q9$ model. For the single-component case, the
equilibrium $f_i^{\rm eq}(\rho,\jj)$ to second order in momentum
is obtained by setting $p/\rho=T_0=1/3$ in Eq.
(\ref{Tequilibrium}). The equilibrium of the mixture is then
described by the populations $f_{{\rm A,B}i}^{\rm eq}=f_i^{\rm
eq}(\rho_{\rm A,B},\jj)$.


The case $\Sc<\tilde{\Sc}$ has  been already considered recently
\cite{KyotoBinary,PREBinary}, and is mentioned here for the sake
of completeness. In this case we consider the momenta of the
components, $\jj_{\rm A,B}$ as the quasi-conserved variables $N$,
and the corresponding quasi-equilibrium is immediately read off
the equilibrium of the single-component fluid; to second order in
momentum it is $f^{*}_{{\rm A,B}i}(\rho_{\rm A,B},\jj_{{\rm
A,B}})=f_i^{\rm eq}(\rho_{\rm A,B},\jj_{{\rm A,B}})$. The
continuous QE model recovers equations of the isothermal
hydrodynamics that is, the advection-diffusion equations for the
densities $\rho_{\rm A,B}$ and the Navier-Stokes equation for the
momentum $\jj$ with $\mu=\tau_1p_0,\ D_{\rm
AB}=\tau_2p_0\frac{X_{\rm A}X_{\rm B}}{m_{\rm AB}}$, where
$p_0=T_0(n_{\rm A}+n_{\rm B})$ is the pressure at the reference
temperature $T_0$. This model recovers
$\Sc=(\tau_1/\tau_2)\tilde{\Sc}$, and according to the hierarchy
of the relaxation times (\ref{hierarchy}), pertains to the fluids
with $\Sc\le\tilde{\Sc}$.


In the opposite case, $\Sc>\tilde{\Sc}$, we choose the stress
tensors of the components, $\bTheta_{\rm A,B}$, as the
quasi-conserved fields $N$ (see Eq. \ref{Pr>}). Corresponding
quasi-equilibria, $f^{*}_{{\rm A,B}i}(\rho_{\rm A,B},\jj,
\bTheta_{\rm A,B})$, are constructed using the triangle entropy
method in the same way as the case $\Pr>\Pr_{\rm LBGK}$.

%

\begin{figure}[htbp]
\centering
  \includegraphics[scale=0.4]{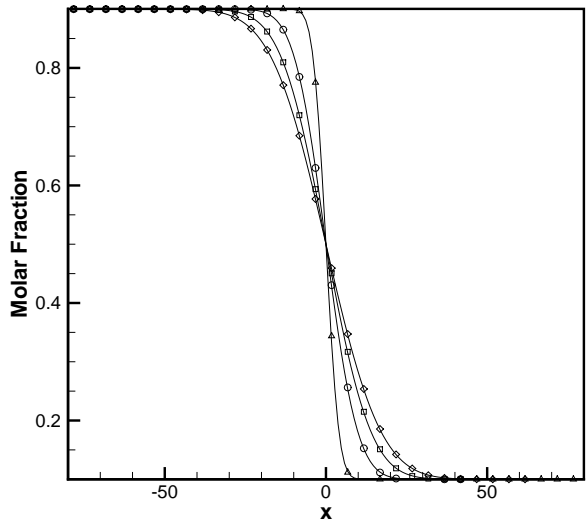}
\caption{Binary diffusion for the case $\Sc>\tilde{\Sc}$ and
$m_B/m_A=500$ at different time steps $t$. Symbol: simulation;
Line: analytical solution. Triangles: $t=500$; circles: $t=3000$;
squares: $t=6000$; diamonds: $t=9000$. }
\label{Fig2}
\end{figure}

The lattice Boltzmann implementation is based on
(\ref{deliverable}) and an interpolation step, as explained in
\cite{KyotoBinary,PREBinary}. In Fig.\ \ref{Fig2}, we present a
simulation of diffusion  of two fluids with a high mass ratio
$m_{\rm B}=500m_{\rm A}$ in a setup where initially a half-space
is filled with the $90-10\%$ mixture and the other half-space -
with the $10-90\%$ mixture.
Agreement with the analytical solution is excellent.

\section{Conclusion}

In conclusion, we suggested a systematic, physically transparent
and realizable approach to constructing lattice Boltzmann models
for hydrodynamic systems. All the models considered herein require
only the choice of the quasi-equilibria appropriate to the
physical context of the problem. Following the same pattern, it is
straightforward to construct kinetic models for bulk viscosity and
chemical reactions.

%
%
%
%


\end{document}